\documentclass[acmsmall]{acmart}

\usepackage{pdflscape}
\AtBeginDocument{%
  \providecommand\BibTeX{{%
    \normalfont B\kern-0.5em{\scshape i\kern-0.25em b}\kern-0.8em\TeX}}}

\usepackage{algorithmic}
\usepackage{graphicx}
\usepackage{textcomp}
\usepackage{xcolor}
\usepackage{listings}
\usepackage[ruled,vlined]{algorithm2e}
\usepackage{tikz}
\usepackage[perpage]{footmisc}

\def\BibTeX{{\rm B\kern-.05em{\sc i\kern-.025em b}\kern-.08em
    T\kern-.1667em\lower.7ex\hbox{E}\kern-.125emX}}
\usepackage{rotating}   
\usepackage{multirow}  
\usepackage{rotating}
\usepackage[english]{babel}
\usepackage[utf8]{inputenc} 
\usepackage[T1]{fontenc}
\usepackage{tabularx}
\usepackage{pdfpages}
\usepackage{balance}
\usepackage{url}
\usepackage{adjustbox}
\usepackage{amsthm}
\newcolumntype{R}[2]{%
    >{\adjustbox{angle=#1,lap=\width-(#2)}\bgroup}%
    l%
    <{\egroup}%
}

\usepackage{lscape}

\usepackage{pdflscape}

\usepackage[most]{tcolorbox}

\usepackage{listings}
\usepackage{color}
\usepackage[labelfont=bf]{caption} 
\usepackage{ragged2e}
\usepackage{lscape} 
\usepackage{subcaption}
\usepackage{pgfplots} \pgfplotsset{compat=1.14}

\usepackage{amsmath}
\usepackage{tcolorbox}
\definecolor{lightGrey}{rgb}{0.9, 0.9, 0.9}

\usepackage{xcolor}

\definecolor{airforceblue}{rgb}{0.36, 0.54, 0.66}
\definecolor{dkgreen}{rgb}{0,0.6,0}
\definecolor{gray}{rgb}{0.5,0.5,0.5}
\definecolor{mauve}{rgb}{0.58,0,0.82}
\definecolor{antiquefuchsia}{rgb}{0.57, 0.36, 0.51}
\definecolor{applegreen}{rgb}{0.55, 0.71, 0.0}
\definecolor{asparagus}{rgb}{0.53, 0.66, 0.42}
\usepackage[graphicx]{realboxes}
\usepackage{color, colortbl} 
 \definecolor{Gray}{gray}{0.9}
  \usepackage{listings,xcolor,tikz}

\usepackage{xspace}
\usepackage{epstopdf} 
\usepackage{amsmath}
\usepackage{tikz}
\usepackage{multirow} 

\newcommand{\tool}{{\it NeuraLint}}

\begin{document}

\title{Automatic Fault Detection for Deep Learning Programs Using Graph Transformations}




\author{Amin Nikanjam}
\authornote{Both authors contributed equally to this research.}
\affiliation{%
 \institution{K. N. Toosi University of Technology}
 \city{Tehran}
 \country{Iran}}
 \affiliation{%
 \institution{SWAT Lab., Polytechnique Montreal}
 \city{Montreal}
 \country{Canada}}
\email{nikanjam@kntu.ac.ir,amin.nikanjam@polymtl.ca}

\author{Houssem Ben Braiek}
\authornotemark[1]
\affiliation{%
  \institution{SWAT Lab., Polytechnique Montreal}
  \city{Montreal}
  \country{Canada}
  }
\email{houssem.ben-braiek@polymtl.ca}

\author{Mohammad Mehdi Morovati}
\affiliation{%
  \institution{SWAT Lab., Polytechnique Montreal}
  \city{Montreal}
  \country{Canada}
  }
\email{mehdi.morovati@polymtl.ca}

\author{Foutse Khomh}
\affiliation{%
 \institution{SWAT Lab., Polytechnique Montreal}
 \city{Montreal}
  \state{Quebec}
 \country{Canada}
 }
\email{foutse.khomh@polymtl.ca}



\renewcommand{\shortauthors}{Nikanjam et al.}

\begin{abstract}
Nowadays, we are witnessing an increasing demand in both corporates and academia for exploiting Deep Learning (DL) 
to solve complex real-world problems. A DL program encodes the network structure of a desirable DL model and the process by which the model learns from the training dataset. Like any software, a DL program can be faulty, which implies substantial challenges of software quality assurance, especially in safety-critical domains. It is therefore crucial to equip DL development teams with efficient fault detection techniques and tools. In this paper, we propose \tool{}, a model-based fault detection approach for DL programs, using meta-modelling and graph transformations. First, we design a meta-model for DL programs that includes their base skeleton and fundamental properties. Then, we construct a graph-based verification process that covers 23 rules defined on top of the meta-model and implemented as graph transformations to detect faults and design inefficiencies in the generated models (i.e., instances of the meta-model). First, the proposed approach is evaluated by finding faults and design inefficiencies in 28 synthesized examples built from common problems reported in the literature. Then \tool{} successfully finds 64 faults and design inefficiencies in 34 real-world DL programs extracted from Stack Overflow posts and GitHub repositories. The results show that \tool{} effectively detects faults and design issues in both synthesized and real-world examples with a recall of 70.5 \% and a precision of 100 \%. Although the proposed meta-model is designed for feedforward neural networks, it can be extended to support other neural network architectures such as recurrent neural networks. Researchers can also expand our set of verification rules to cover more types of issues in DL programs.%
\end{abstract}



\maketitle

\section{Introduction}
\begin{figure}[h]
\centering
\includegraphics[scale=0.9]{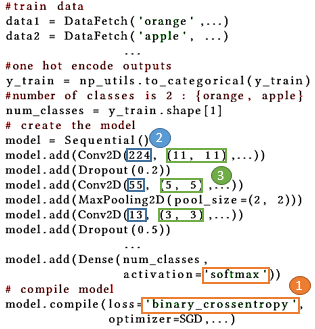}
\vspace{-5pt}
\caption{Simplified example DL program from SO\_44322611.}
\label{fig:code}
\vspace{-10pt}
\end{figure}

Many developers, entrepreneurs, and researchers are showing an increasing enthusiasm in developing and using Deep Learning (DL) in a variety of domains. The core of DL programs is deep neural networks. After building a desirable DL model, it should be trained by executing a learning algorithm on a dataset. Easy-to-use libraries such as TensorFlow or Keras have been developed to simplify the 
development process. However, leveraging these libraries to implement a training program is still challenging, in particular for developers who are not experts in machine learning and neural networks. Like any software, a DL program often contains design issues and bugs. For example let's consider the program from Figure~\ref{fig:code} which is extracted from Stack Overflow (SO) post \#44322611 and is reported to have a low accuracy. 

This program which implements a Convolutional Neural Network (CNN) has three issues. 
The first issue (i.e., \raisebox{.5pt}{\textcircled{\raisebox{-.9pt} {1}}}) is a bug due to the incompatibility between the \textit{softmax} as output activation and \textit{binary\_crossentropy} as loss function. In fact, developers should have chosen \textit{categorical\_crossentropy} because it works with one-hot encoding labels and softmax activation to solve multi-label classification problems (including binary labels problems)\footnote{We refer the reader to \cite{loss_and_acts} for more practical use cases about how to choose the last layer activation and loss function when using the Keras Library.}. This kind of API misuse error can be identified by 
verifying the consistency of the involved mathematical operations. For instance, error \raisebox{.5pt}{\textcircled{\raisebox{-.9pt} {1}}} induces an inconsistency between the loss function and the last layer activation. 
The other two errors are \raisebox{.5pt}{\textcircled{\raisebox{-.9pt} {2}}} decreasing filters count ${224 > 55 > 13}$ and \raisebox{.5pt}{\textcircled{\raisebox{-.9pt} {3}}} decreasing filtering spatial size ${(11,11) > (5,5) > (3,3)}$, which represent poor structural CNN choices that violate the common design patterns of effective and optimal CNN architectures~\cite{CNN_design_patterns, CNN_principles}. These structural errors are often detected through manual code reviews which is time consuming. A static code analysis using automated tools can significantly speed up this process. 
In this paper, we examine common structural errors and design inefficiencies occurring in DL programs and propose \tool{}, a model-based verification approach for their detection. To design \tool{}, we first propose a meta-model for DL programs that includes their base skeleton and fundamental properties. This meta-model captures their essential properties independent of available DL libraries. Considering the proposed meta-model, we specify for each fault or design issue, a verification rule that can be used to detect its occurrence. Finally, we propose a checking process to verify models of DL programs that are conforming to the meta-model. We employ graph transformations to implement \tool{}. We present a type graph for the meta-model and graph transformations for the verification rules. 

We evaluate our approach \tool{} by finding various types of faults and design issues in 28 synthesized examples built from common problems reported in the literature~\cite{DL_faults} and 34 real-world DL programs extracted from GitHub repositories and SO posts. The results show that \tool{} effectively detects faults and design issues in both synthesized and real-world examples. 
This paper makes the following contributions.
\begin{itemize}
\item We propose a meta-model for DL programs;
\item We describe 23 common errors and poor design practices of DL training programs and provide rules for their detection;
\item We propose a model-based verification approach for DL programs, using meta-modelling and graph transformation rules. 
\item We provide a concrete implementation of the approach as a tool that DL developers can use to detect errors and design issues in their DL training programs. 
\end{itemize}
\textbf{The remainder of this paper is organised as follows.} Section~\ref{background} provides background information about graph transformation systems, and deep neural networks. Section \ref{meta-model} presents our meta-modeling of DL programs. Section \ref{rules} introduces the studied issues and rules for their detection. Section~\ref{approach} presents our proposed approach \tool{}. Section~\ref{evaluation} reports the empirical evaluation of \tool{}. Section~\ref{relatedWork} discusses the related literature. Finally, Section~\ref{conclusion} concludes the paper and discusses future work.

\section{Background}\label{background}
In this section, we introduce background knowledge about graph transformation systems, and deep neural networks.
\subsection{Graph Transformation Systems}
Graph transformation system (GTS) \cite{heckel2006graph} (also called graph grammar) is a formal language for the specification of software systems, in particular those 
with dynamic structures. The definition of an attributed GTS consists of a triplet \textit{(TG, HG, R)} in which \textit{TG} is a type graph, \textit{HG} is a host graph, and \textit{R} is a set of rules for graph transformation. \textit{TG} is defined by four components, \textit{$TG=(TG_N, TG_E, src, trg)$}. \textit{$TG_N$} and \textit{$TG_E$} includes all node types and edge types respectively. \textit{src} and \textit{trg} are two functions \textit{$src: TG_E \rightarrow TG_N$} and \textit{$trg: TG_E \rightarrow TG_N$}, that determine the source/destination nodes of an edge, respectively. The initial configuration of a system specified by GTS is presented by the host graph which is an instance of the type graph. Therefore, each component of the host graph, node or edge, must have a component type in the type graph. A host graph \textit{HG} may instantiate from a type graph \textit{TG} using a graph morphism function \textit{$type_G: HG \rightarrow TG$}, in which the components of \textit{HG} are instantiated from \textit{TG}. Other configurations or states of a system are generated by successive applications of transformation rules on the host graph. A transformation rule \textit{r} in \textit{R} is defined by a triplet \textit{$(LHS_r, RHS_r, NAC_r)$} in which \textit{$LHS_r$} (left-hand side) represents the preconditions of the rule whereas \textit{$RHS_r$} (right-hand side) describes the postconditions. Moreover, there may be a Negative Application Condition (NAC) for the rule \textit{r}, meaning that the rule \textit{r} can be applied only when \textit{$NAC_r$} does not exist in the host graph. By applying the rule \textit{r} to the host graph \textit{HG}, which is an instance model of the meta-model or type graph, a matching of the \textit{$LHS_r$} in \textit{HG} is replaced by \textit{$RHS_r$}. Formally, a graph morphism exists between \textit{$LHS_r$} and the instance model \textit{HG}. The application of a rule is performed in four steps: (1) find a matching of \textit{$LHS_r$} in \textit{HG}, (2) check \textit{$NAC_r$} that forbid the presence of certain nodes and edges, (3) remove a part of the host graph that can be mapped to \textit{$LHS_r$} but not to \textit{$RHS_r$}, and (4) add new nodes and edges that can be mapped to the \textit{$RHS_r$} but not to the \textit{$LHS_r$}.
\subsection{Deep Learning Software Development}
In this section, we first introduce the architectures of Deep Neural Networks (DNNs) supported by our meta-model and verification approach. Then, we present the state-of-practice regarding the software implementation of these DNNs. 
\subsubsection{Feedforward Neural Network (FNN)}
FNN~\cite{DL_ebook_2016} is the principal neural network architecture used for solving classification and regression problems, where the task is to learn a mapping function capable of converting input data to a target output. FNN consists of many, and sometimes diverse, sequences of layers of computational units. These computational layers are trained to extract features hierarchically (i.e., starting from low-level to high-level features), then, detect discriminative and informative patterns, which serve the FNN to derive either the class label (classification problems) or continuous outcome (regression problems). It is called feedforward because the information flows in a forward manner from the input layer, through the hidden layers (if any) and to the output layer, e.g., a class probability or a predicted real value.
\paragraph{\textbf{Computational Layers.}} FNN layers are constructions of neurons, where each performs:\textit{(i) A linear calculation} that consists of a weighted sum of all the signals from the previous layer with addition of a constant (i.e., bias); \textit{(ii) A non-linear activation} that consists of a gate function, filtering out the computed quantities, to derive the input signals for the neurons of the next layer. The initialization of layers' parameters (i.e., weights and biases) and the choice of activation functions represent essential parts of the FNN design that have a crucial impact on the performance of the training procedure \cite{hayou2019impact}.
\paragraph{\textbf{Parameters Training.}} Starting from a random initialization, the training process consists in updating iteratively the model parameters, towards minimizing the loss of DNN's predictions as regards to the training labeled data. Indeed, a loss/cost function is defined to estimate the average distance between predicted and actual outcomes. The training pass over FNN's parameters is performed by backpropagation that makes the loss gradients flow over the neurons defined in the network but in the opposite direction, in order to update the parameters of each layer. The update quantities are mostly approximated using a first-order optimization algorithm such as SGD, Momentum, and Adam. Thus, the training procedure resides in fulfilling several training passes on \textit{i.i.d} batches sampled from training data (independent and identically distributed) until reaching a local or the global minimum of the loss. Commonly, the best-fitted FNN is found after multiple epochs (i.e., passes over all the training data).
\paragraph{\textbf{Regularization Methods.}}
The regularization is required to improve the convergence and generalizability of the above-mentioned parameters training procedure. For DNNs, many regularization techniques have been proposed and the most used ones are dropout and batch-normalisation (batchnorm). Dropout~\cite{dropout} masks at every training iteration a random subset of units (i.e., nullify them). The stochasticity injected into the inference calculation, only during the training, prevents the co-adaptation of feature detectors and encourages the DNN to learn robust patterns against partially-hidden information. Batchnorm~\cite{batchnorm} acts differently on activations by normalizing their values using statistics (i.e., mean and variance) of the current batch of data during the training. 
During the testing, it updates internally, the population statistics of all batches for each level of activations in order to switch to normalizing against population, rather than batch, statistics. This normalization of intermediary input data has shown its effectiveness in smoothing the loss landscape, which ensures faster and safer training convergence with high potential to escape weak local minima.
\paragraph{\textbf{Dense and Convolutional Architectures.}}
The basic FNN architecture consists of stacking dense layers, where all the neurons of two consecutive layers are fully-connected. In this paper, we consider convolutional architectures which represent a particular type of FNN designed for multi-dimensional input data, such as 2D images and audio spectrograms, or 3D videos. The benefit of CNNs lies in their ability to take into account the spatial information in their feature extraction process. To do that, CNNs stack, earlier, two specialized layers: convolutional layer and pooling layer.
\textit{(1) Convolutional layer} applies spatial filters over the input data and each filter’s weights are learned to detect relevant features supporting the network’s task. Thus, it yields a feature map for each learned filter, where each unit is connected to a local region (i.e., size of spatial filtering window) in its previous layer’s feature maps and \textit{(2) pooling layer} performs spatial pooling over the computed feature map to reduce its dimensionality and retain the most relevant information. The spatial pooling can be either average or max aggregation that computes, respectively, the average or max of all the units in the specified spatial window. Conventionally, a CNN architecture incorporates a bundle of convolutional layers with increasing filter count and separated by pooling layers to shrink gradually the feature map area. Hence, the extracted feature space tends to become deeper and narrower throughout the network until it becomes ready to be flattened and fed to the dense layers in charge of mapping the features into the target output.
\subsubsection{DL Software Development}
DL libraries encode the DNN as a Directed Acyclic Graph (DAG), where nodes and edges represent, respectively, operations and data paths. This computational graph represents a high-level of abstractions to organize the components and their relations. Thus, the internal calculations can be decentralized where each component performs its local computation and communicates the result via data paths to its connected components. Besides, the DAG abstraction allows the DL library's graph compiler to optimize the computations’ execution for a target hardware architecture.\\
Leveraging DL libraries to implement a training program for a designed DNN is not straightforward and it can be error-prone.
DL libraries often have to trade off between the coverage of novel DL functionalities and the ease of rapid implementation and extension of DNN software prototypes. As a compromise solution, they uniformly include, for each newly-implemented DL functionality, a bundle of automated steps and default settings following its common usage trends in the community. This enables quick prototyping of regular DNNs while keeping the flexibility to try other configurations with the tweakable setting options available for every provided DL routine. As a consequence, DL developers should be aware of the intricacies of these DL libraries to choose the appropriate configurations and avoid breaking their implicit assumptions in regard to the usage of their built-in routines. 
\section{Meta-modeling DL Programs}\label{meta-model}
With the proliferation of libraries supporting the development of DL programs, a fundamental question emerges: is there any generic representation of DL programs that is independent from these libraries? In other terms, can we define a meta-model of DL programs and how can we model a DL program? Answering this question would pave the way for the application of model-driven engineering techniques to the detection of errors in DL programs. In \cite{hartmann2019meta}, researchers proposed a meta-model for meta-learning. They presented an overview of the meta-learning concepts –on a meta-modelling level– with possible variabilities and discussed how their meta-model could be integrated into existing modelling frameworks and tools. However, while their meta-model includes "Learning Block", "Learning Algorithm", "Optimizer" and "Hyperparameters", no further details like specifications of learning algorithms or blocks are presented and they did not explore the possibility of identifying errors in machine learning models. In this section, we present a particular meta-model for DL programs and our approach for meta-modeling of such programs to perform static analysis of DL programs. We describe possible variabilities of the meta-model and how concrete DL programs can be generated from it. We think that a generic meta-model for DL programs can significantly facilitate the use of DL in various applications and would be helpful for understanding DL programs written by developers using third-party DL libraries. In fact, model-driven engineering is a perfect tool to make this idea come to life and ease the process of developing and debugging DL programs.
\begin{figure}
  \includegraphics[width=\textwidth]{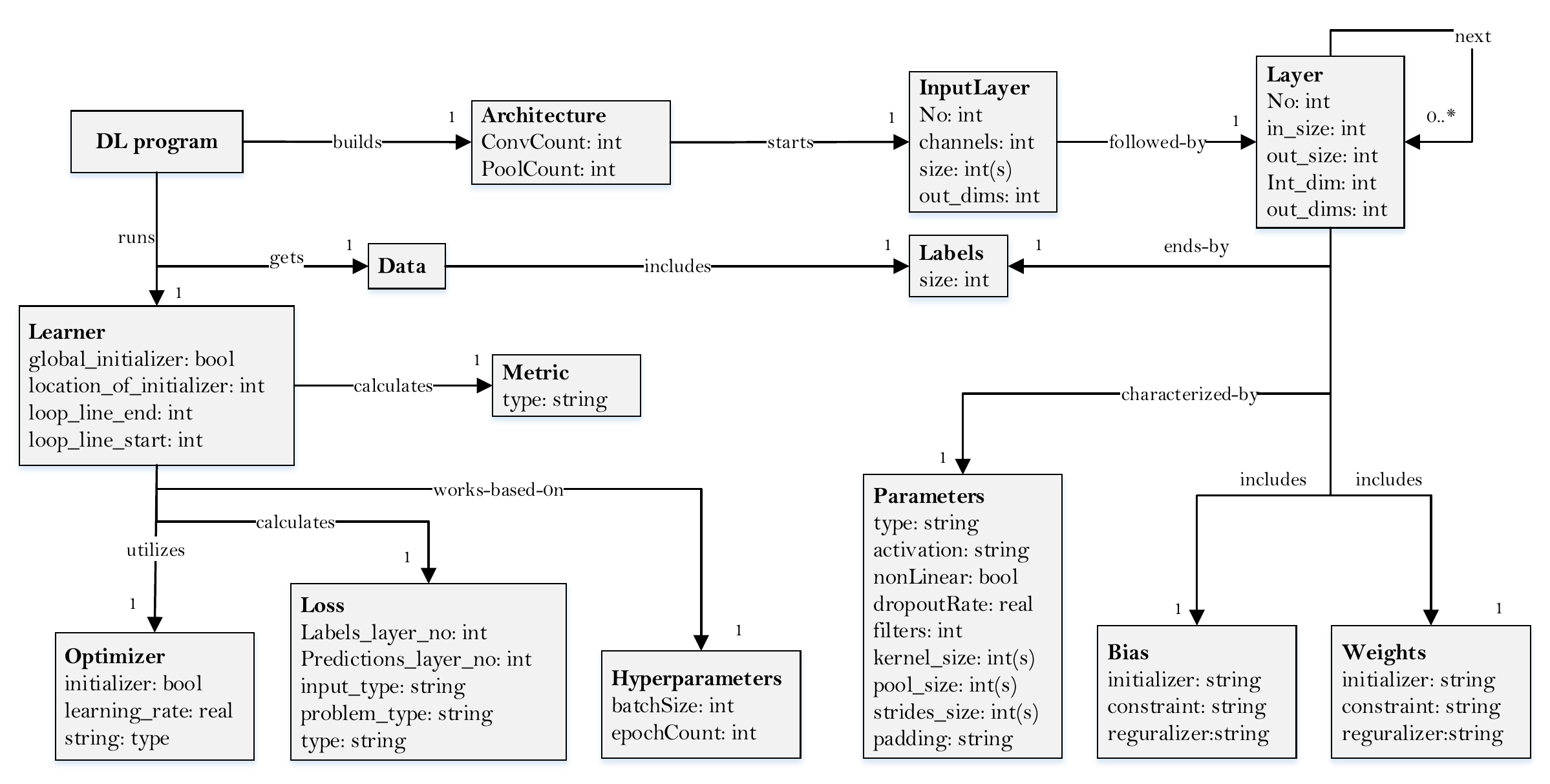}
  \vspace{-15pt}
  \caption{The proposed meta-model (type graph) for DL programs.}
  \label{fig:typegraph}
  \vspace{-10pt}
\end{figure}
\subsection{A Meta-Model for Deep Learning Programs}
A DL program has different components. The core of each DL program is a DNN. For the sake of simplicity, we only consider the feedforward multilayer perceptron (MLP) architecture. Like other computational models, DNN attempts to find a mathematical mapping from the input into the output during a learning phase. Usually, a set of inputs and desired outputs (or targets) is provided for learning, which is called Dataset. Therefore, our meta-model includes three main parts: Architecture of DNN, Learner, and Data. Since we have used GTS for modeling DL programs, our proposed meta-model is represented by a type graph. The proposed type graph is illustrated in Figure \ref{fig:typegraph}. The node representing the \textbf{DL program} has three edges to \textbf{Architecture}, \textbf{Learner} and \textbf{Data} nodes indicating its main components. In the following, we describe the meta-model in detail. It should be noted that our aim of meta-modeling is the detection of faults in DL programs; therefore the most relevant components have been incorporated into the meta-model.

\subsubsection{Architecture of deep neural network}
An architecture starts with the input layer, continues with some hidden layers and ends with the output layer. We have considered a distinctive node for the \textbf{InputLayer} because of its importance but all other successive layers are modelled as \textbf{Layer}. Each layer has a \textbf{size} indicating the number of neurons in that layer. There are specific properties among nodes that are modelled as edges. For example, \textbf{Architecture} starts by \textbf{Input Layer}, \textbf{Input Layer} is followed by other \textbf{Layer}s, each \textbf{Layer} may have next layers and each \textbf{Layer} has a \textbf{Type} as an attribute. There are different types for a layer in DL, e.g., dense, 1D and 2D convolution, pooling or data processing layers like flatten. There may be other attributes for a layer like \textbf{Bias}, \textbf{Weights}. An architecture ends with \textbf{Labels}, the desired outputs of DNN that are used to calculate the error of the network in \textbf{Loss} function. Actually, \textbf{Labels} is a part of \textbf{Data} associated with the DL program.

\subsubsection{Learning algorithm}
A DL program normally employs a learning algorithm, \textbf{Learner} in our meta-model, to learn the mapping from inputs to outputs. A \textbf{Loss} function is used to calculate the error of a neural network in matching the target (desired) and output value during training. The goal of learning is minimizing the loss by modification of the network's parameters (weights) by an \textbf{Optimizer}, e.g., Adam or stochastic gradient descent (SGD). The overall performance of the network is measured using a \textbf{Metric}. Moreover, there are some \textbf{Hyperparameters} like the number of epochs or batch size.

\subsubsection{Data}
This node contains \textbf{Labels}, meta-data, features, and related information about the data set, like shuffling and batching.

\subsection{DL programs modeling}
A model of a DL program includes components that form its source code. There are two ways to build a model of DL programs: configure an arbitrary model directly or transform a DL program to a model. One may design a model for a DL program that conforms to the proposed meta-model by configuring each component of the meta-model. Starting from an empty model, layers are added one by one to \textbf{Architecture}; making a chain of layers that starts by \textbf{InputLayer} node, follows by other \textbf{Layer} and ends with \textbf{output}. Each layer is configured separately to set \textbf{Parameters}, \textbf{Weights}, etc, and once a layer is configured completely, the model will proceed to the next layer. Other components of DL programs like \textbf{Learner} and \textbf{Data} are configured respectively. This process is similar to what a developer does when developing a DL program using popular DL frameworks. Therefore, the meta-model and resulting models would be realistic from a practitioner point-of-view and sufficiently flexible in representing plenty of DL programs.

On the other hand, a model could be configured according to a DL program that has already been developed by a programmer. The source code of a DL program is converted to a model, which is an attributed graph. Dedicated convertors are programmed in \tool{} to convert a DL program written by different DL libraries to its model. The source code of a program is parsed to extract relevant information that is necessary to configure the model. The meta-model is generic enough to be independent of any specific DL library. Hence, we can have a model of a DL program that conforms to the meta-model; making possible further investigations on the model, such as verification. Apart from the work and analysis that are presented in the rest of this paper, we believe that this meta-model can be very useful to understand DL programs written by third parties. It will be helpful in understanding the development activities of DL practitioners; the way they write DL programs and the type of faults that they experience.

\section{Model-based Verification Rules}\label{rules}
In this section, we present the proposed rules for detecting faults in DL programs. First, we report the adopted methodology for extracting rules. The rules are then described in Subsection \ref{ruless}. Afterward, we present a discussion on the application scope of rules. Finally, we describe the approach adopted to implement the rules using graph transformations.

\begin{figure}[t]
  \includegraphics[width=0.85\textwidth]{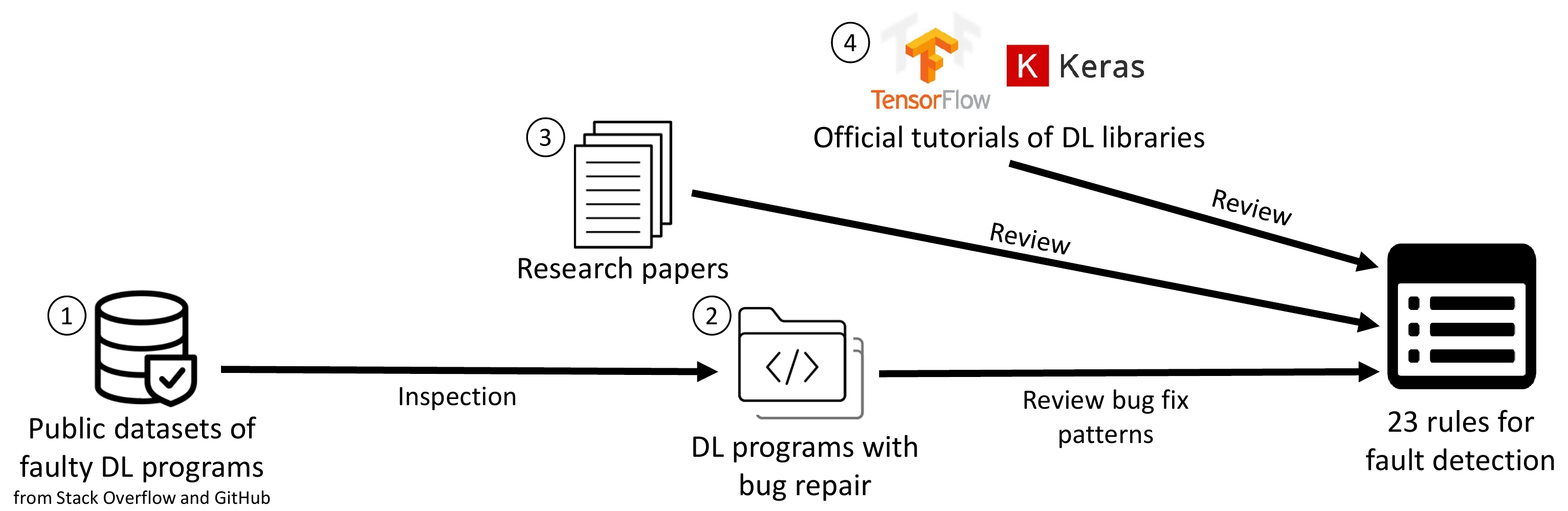}
  \caption{The adopted methodology for extracting rules from different sources.}
  \label{fig:methodology}
\end{figure}

\subsection{Methodology}
Figure \ref{fig:methodology} illustrates the adopted methodology for extracting the rules. We have explored three main sources: datasets of buggy DL programs (including bug repairs), relevant research papers and official DL libraries’ tutorials. In the first step, we manually inspected the labeled bugs in the public datasets of buggy DL programs released by former research studies \cite{DL_bugs_1, DL_bugs_2, DL_faults} collected from SO and GitHub with the objective of discovering finer root causes of bugs. Zhang et al. \cite{DL_bugs_1} published the first empirical study on real-world DL bugs occurring in Tensorflow-based software systems including their high-level root causes and symptoms. Then, Islam et al. \cite{DL_bugs_2} extended the investigated cases to include DL software systems written using other competitive DL libraries such as Pytorch and Caffe, and studied the relationship and the evolution of different DL bug types. Last, Humbatova et al. \cite{DL_faults} refined the former bug investigation \cite{DL_bugs_1, DL_bugs_2} into a taxonomy of real faults that occur in DL software systems. In the second step, we have inspected bug fixes suggested in accepted answers of SO posts and fix patterns adopted in GitHub samples to identify patterns followed to fix the reported bugs.\\
We reviewed research studies on DL bugs \cite{DL_bugs_1, DL_bugs_2, DL_faults} and fundamental DL design principles \cite{glorot2010understanding, nwankpa2018activation, CNN_design_patterns, CNN_principles} in the third step. Regarding the former, the aim was finding rules for validating the correctness of model structure and configuration choices through the DL program’s model drawn from the code. We build on these previous works to specify rules that can be used to detect occurrences of different types of issues in DL programs and validate the conformity of the DNN design to common patterns through static code inspection. However, most of the DNN design patterns and principles have been deduced from state-of-the-art CNN architectures \cite{systematic_CNNs, practical_CNNs} that have shown their effectiveness on public computer vision datasets and competitions such as ImageNet classification \cite{deng2009imagenet} or COCO object detection \cite{lin2014microsoft}. Thus, we aim to report warnings to the user whenever a poor design choice is spotted with respect to these empirical research studies on DNN design principles. This would likely steer the user to redesign his model in order to avoid the performance degradation either at the training or at the inference mode. Finally, to support practitioners in their DL program debugging, we also proceed with a dual analysis over the commonly reported APIM bugs and the official DL libraries’ tutorials.\\
In the end, we have come up with 23 rules for detecting bugs and issues in DL programs. The rules are organized into different high-level root causes as initially introduced in \cite{DL_bugs_1}, namely Incorrect Model Parameter or Structure (IPS), Unaligned Tensor (UT), API Misuse (APIM), and Structure Inefficiency (SI).  

\subsection{Rules}\label{ruless}
\subsubsection{Incorrect Model Parameter or Structure (IPS)}
IPS bugs are related to modeling faults that arise from either an inadequate model parameter like learning rate or an incorrect model structure like missing or redundant layers \cite{DL_bugs_1}. The major symptom of IPS bugs is anomalous training behaviors leading to low effectiveness such as low precision and a huge loss.\\ \\
\textbf{Rule 1: Asymmetric Units Initialization.} \textit{The initialization of weights should not be constant to break the symmetry between neurons} \cite{glorot2010understanding}. For instance, a common mistake is to start with null weights, which eliminates asymmetry between the neurons (i.e., all the neurons would output the same value, and then, would receive the same gradients). \\
\textbf{Rule 2: Null Biases Initialization.} \textit{The initialization of biases is preferred to be zeros} \cite{glorot2010understanding}. It is a common practice to expect that the outputs could be totally explained by the input features. Indeed, no custom initial bias provided a consistent improvement, but it may weaken the learning.\\ 
\textbf{Rule 3: Non-Linear Activation Requirement.} \textit{Activations for learning layers (i.e., convolution and fully-connected layer) should be a non-linear function.} This key attribute is needed to enhance the ability of DNN to model highly nonlinear mappings and draw complex shape decision boundaries \cite{nwankpa2018activation}.\\ 
\textbf{Rule 4: Unnecessary Activation Removal.} \textit{Multiple and redundant connected activations are not allowed.} Since all activation functions are designed to transform real values into a restricted interval \cite{nwankpa2018activation}, successive activations applied to the same features can make their last activation unable to produce its full output range.\\ 
\textbf{Rule 5: Class Probability Conversion.} \textit{A last layer activation is required to transform the logits into probabilities for classification problems.} In detail, sigmoid ($\sigma(z) = \frac{1}{1 + e^{-z}}$) and softmax ($\sigma(z)_i = \frac{e^{-z_i}}{\sum_{j=1}^Le^{-z_j}}$, for $i=1,...,L$ and $z \in \mathbb{R}^L$) are, respectively, needed to normalize outputs with single unit ($z \in \mathbb{R}$) and multiple units ($z \in \mathbb{R}^L$). A common mistake is to use softmax instead of sigmoid for binary classification without one-hot encoding beforehand, which totally obstructs the learning because the outcome is always equal to $1 = e^{-z_1}/e^{-z_1}$, $z\in\mathbb{R}^1$. \\ \\
Indeed, the above-mentioned rules show that static DL code analysis can help detect earlier structural bugs and misconfigurations, however, incorrect parameters like learning rate or neural network size (width and depth) could be identified through empirical evaluation of the model on the underlying data.
\subsubsection{Unaligned Tensor (UT)}
The computational units in a DNN graph are mostly tensor-based operations, where each one receives and returns tensors (i.e., multi-dimensional arrays). Their connections can hide issues related to the compatibility of tensors’ shapes. DL developers often fail to express and manipulate the shapes of tensors properly [39] because DL libraries mask all the algebra computations and dynamic shapes’ inference details. A bug triggered during the DNN graph construction when the shapes of one operation’s tensors do not match is called an Unaligned Tensor (UT) \cite{DL_bugs_1}. The major symptom of UT bugs is runtime errors because the underlying tensor-based operation could not run on two incompatible tensors. However, the dynamic shape inference included in most DL libraries often makes the exception of incompatible shapes triggering far from its localization in the DL code; so the error message can be misleading. In the following, we describe various DNN layers’ connectivity and configuration rules that can be checked on the DL model to identify the UT bug type and localization.\\ \\
\textbf{Rule 6: Consecutive Layers Compatibility.} \textit{A processing layer that operates on a N-dimensional tensor, should receive a valid input tensor with exactly N-dimensional shape.} For instance, a Conv2D layer works on $4$-D tensors, i.e., $[samples, height, width, channels]$, but a Dense layer works on $2$-D, i.e., $[samples, units]$, which means a reshape layer is needed to flatten the convolutional feature space before starting the dense layers’ inference.\\
\textbf{Rule 7: Spatial Size Agreement.} \textit{A processing layer should receive sufficient-sized feature space to perform its spatial filtering or pooling.} For instance, $2$-D processing layers like Conv2D and MaxPooling2D require a size of feature space greater or equal to their local window size, i.e., ($window\_height \leq input\_height$) and ($window\_width \leq input\_width$).\\
\textbf{Rule 8: Reshaped Data Retention.} \textit{A reshape layer should preserve the total data elements.} More specifically, we verify that the product of original tensor dimensions equals the product of reshaped tensor dimensions.\\
\textbf{Rule 9: Separate Item Preservation.} \textit{A reshape layer should never alter the size of elements (i.e., first dimension).} Otherwise, the reshape would provoke an overlap between data items (i.e., points in the feature space), and as a consequence, invalidate the following layers designed to process each data item, independently. 
\subsubsection{API Misuse (APIM)} 
APIM bugs are the ones introduced by practitioners who misunderstand some essential assumptions made by the used DL APIs \cite{DL_bugs_1}. Indeed, most DL libraries encode the DNN as an acyclic computational graph where the edges are tensors and the nodes correspond to operations. The operations include all the supported computational units that form the linear computations, activations, gradient estimations, etc. Programmatically, practitioners describe their designed DL program by inserting and configuring built-in DL routines, and connecting them by putting the outputs of one operation as inputs to another. When these routines are added without fulfilling their usage conditions or without context alignment, the DL program would not reflect the designed DL model or cannot be successfully executed by the DL core framework, which leads, respectively, to low effectiveness or runtime exceptions. Below, we detail the verification rules that should be executed on the generated static analysis-based graph model to confirm the existence of essential DL program’s components and their consistency with API assumptions and recognized application context.\\ \\
\textbf{Rule 10: Valid Loss Linkage.} \textit{The loss should be correctly defined and connected to the last layer activation in accordance with its input conditions (i.e., shape and type).} For instance, the input type for cross-entropy based losses could be either logits or probabilities. Indeed, numerically stable implementations regarding the cross-entropy based losses require merging both loss and last activation functions together to rewrite the join formula carefully without any risk of $\log(0)$ or $\exp(\infty)$. However, ignoring this difference between theoretical loss functions and their numerically-stable implementations gives rise to a common mistake in the development of DNN programs, as passing activated output to this logit-based loss would cause redundant activations.\\ 
\textbf{Rule 11: Valid Optimizer Linkage.} \textit{The optimizer should be correctly defined and connected to the computational graph.} Depending on the DL library, it could be either connected to the loss (e.g., TensorFlow) or the learnable parameters (e.g., Pytorch).\\
\textbf{Rule 12: Single Global Initialization.} \textit{The learnable parameters should be totally initialized once at the beginning of the training.} For some DL libraries (e.g., TensorFlow), this mandatory condition should be carried out by the developer.\\ 
\textbf{Rule 13: Zero Gradients Reset.} \textit{The gradients should be re-initialized after each training iteration.} This clears old gradients from the last step; otherwise accumulating the gradients hinders the optimization process. Some DL libraries (e.g., Pytorch) delegate this necessary reset step to their users. \\
\textbf{Rule 14: Iterative Training Procedure.} \textit{The loss minimization problem should be solved iteratively with continuous update of parameters.} Depending on the granularity level of the API used, it could be a native loop of optimization routine calls or a single call of a configurable fit function.

\subsubsection{Structure Inefficiency (SI)}
SI issues reflect a misconfiguration in the DNN design and its structure that leads likely to performance problems, contrary to IPS bugs that leads to functional incorrectness \cite{DL_bugs_1}. SI issues may result in performance inefficiencies (like long time of model training/inference) or poor predictions (like low classification accuracy). As an example, large feature-maps, especially in the early layers, provide more valuable information for the CNN to utilize and improve its discriminative power. Therefore, it is crucial to avoid prematurely down-sampling and excessive appliances of pooling. Otherwise, the model will lose some information extracted in early layers resulting in poor performance. Since the best trained model cannot guarantee 100\% of accuracy, it is challenging to detect design issues by assessing the performance of the obtained models. Indeed, some misconfigurations and poor design choices may definitely introduce inefficiencies on the internal functioning of the DNN or one of its components, which can hinder the expressiveness of mapping functions, memory and compute consumption. For example, when increasing the depth of a DNN, it is important to control both the model size and the computational cost (regarding the specific task); otherwise, stacking a high number of layers can worsen the performance.\\ \\
\textbf{Rule 15: Effective Neurons Suspension.} \textit{The dropout layer must be placed after the maximum pooling layer to be more effective.} Considering the case studies with max-pooling layers~\cite{dropout}, the dropout has been applied on the pooled feature maps, which becomes a heuristic followed by the state-of-the-art CNN architectures~\cite{systematic_CNNs, practical_CNNs}. The intuitive explanation is that dropping out the activation before the pooling could have no effect except in cases where the masked units correspond to maximums within input pooling windows because the max-pooling would keep only these maximums as inputs for next layers.\\
\textbf{Rule 16: Useless Bias Removal.} \textit{A learning layer should no longer include a bias when it is followed by batchnorm.} Batchnorm applies, after the normalization, a linear transformation to scale and shift the normalized activations $\hat{a_i} = \alpha a_i + \beta$, where $\alpha$ and $\beta$ are learnable parameters. This allows DNN to compensate for any loss of information by the value distortions in order to preserve its expressive power. Since, batchnorm already adds a $\beta$ term fulfilling the same role of bias, so ``its effect will be canceled''~\cite{batchnorm} in the presence of a bias.\\
\textbf{Rule 17: Representative Statistics Estimation.} \textit{Batchnorm layer should be before the dropout.} Otherwise, batchnorm computes non-representative global statistics (i.e., moving average and moving variance) on the dropped outputs of the layer. Li et al. \cite{disharmony_dropout_batchnorm} discussed the reason behind this disharmony between dropout and batchnorm and showed experimental results reinforcing their explanation.\\
\textbf{Rule 18: Pyramid-shaped Construction.} \textit{The area of feature maps and the width of fully-connected units should be progressively decreasing over the layers.} It has been shown~\cite{pyramidal_DNN} that the progressive size reduction of activations implicitly forces the neural network to find and learn more robust features. Hence, it significantly improves its predictions, since the network decisions are based on more discriminative and less noisy features.\\
\textbf{Rule 19: Maximum Pooling Domination.} \textit{Max-pooling is the preferred down-sampling strategy.} In fact, down-sampling~\cite{strided_conv} can be done by max- or average-pooling or strided convolution (strides greater than $1$). Nevertheless, max-pooling operation has been shown~\cite{MaxPooling_Sup} to be extremely superior for capturing invariances in data with spatial information, compared to other downsampling operations.\\
\textbf{Rule 20: Gradual Feature Expansion.} \textit{The number of feature maps should be gradually expanded while the feature map area is retracted.} The growth of feature maps count is recommended~\cite{depth_comp} to compensate for the loss of representational expressiveness caused by the continuous decreasing of the spatial resolution of the learned feature maps. Throughout the layers, the feature space becomes synchronously narrower and deeper until it gets ready to be flatten and fed as input vector to the dense layers.\\
\textbf{Rule 21: Local Correlation Preservation.} \textit{The local window size for spatial filtering should generally increase or stay the same throughout the convolutional layers.} It makes sense that by using CNNs, the locality of information is crucial for performing the task. Thus, it is important to preserve locality throughout CNN to guarantee its success in detecting various features and relations between them~\cite{lecun2015deep}. Furthermore, early convolutional layers learn lower level features while deeper ones learn more high-level and domain specific concepts.
It is recommended~\cite{VGGNet, szegedy2016rethinking} to start with small spatial filtering to collect much local information and then gradually increase it to represent more compound information.
\\
\textbf{Rule 22: Maximum Information Utilization.} \textit{Deep CNN should not apply pooling after every convolution.} For instance, we use, as approximations, the minimum of 10 layers to consider a CNN deep and $1/3$ as threshold for the proportion of pooling layers with respect to the total of convolutional layers (convolution + pooling) to pinpoint a high amount of pooling. In fact, it has been shown~\cite{he2015convolutional, szegedy2016rethinking, iandola2016squeezenet} that larger feature-maps, especially in the early layers, provide more valuable information for the CNN to utilize and improve its discriminative power. Therefore, it is crucial to avoid prematurely down-sampling and excessive appliance of pooling.\\
\textbf{Rule 23: Strive for Symmetry and Homogeneity.} \textit{Deep CNN should favor blocks of $2$, $3$ or even $4$ homogeneous convolutional layers with similar characteristics.} Indeed, going deeper does not refer to simply maintaining stacking a series of convolution and pooling layers. Advanced CNN architectures~\cite{krizhevsky2012imagenet, he2016deep, iandola2014densenet} have shown the benefit of having several homogeneous groups of layers, where each one is specialized to achieve a particular goal. Indeed, building blocks of convolutional layers with similar characteristics (i.e., the same number of feature maps and feature map sizes) increases the homogeneity and the structure symmetry within the CNN. Hence, larger kernels can be replaced into a cascade of smaller ones, which enhances the nonlinearity and yields better accuracy~\cite{VGGNet}. For instance, one $5\times5$ can be replaced by two $3\times3$ or four $2\times2$ kernels. Moreover, spatial filtering with reduced size decreases massively the computation power requirement because recent NVIDIA cuDNN library  (version 5.x or higher) is not optimized for larger kernels such as $5\times5$ and $7\times7$, whereas CNN~\cite{VGGNet} with entirely $3\times3$ filters achieved a substantial boost in cuDNN performance.
\subsection{Application scope} \label{scope}
The rules are defined to support the debugging of DL programs through static analysis-based graph models. On the first hand, we have been limited to the information that could be parsed from the source code of a DL program. For instance, there are some model parameters that should be experimentally tested to assess their adequacy for the underlying problem, particularly for IPS bugs. Thus, the bugs related to data (type, format, and preprocessing steps) and hardware issues (GPU configuration and required memory) are excluded from the rules and debugging scope because the information needed to diagnose the issue and identify those bugs are mostly out of the static DL code scope. In fact, these types of bugs could be better detected using Python and GPU firmware native debugging tools that help inspect step by step the executed statements at runtime. On the other hand, we have defined a high-level meta-model that could be instantiated to represent any DL program; so, bugs and intricacies that are related to specific DL libraries or APIs are discarded from the verification routines. Referring to the identified high-level root causes of DL bugs \cite{DL_bugs_1}, we did not consider API Change (APIC), which reflects anomalies by a DL program upon a new release of the used library and Confusion with Computation Model (CCM), which includes bugs arising from misunderstanding the DL library computation model such as DAG and deferred execution of Tensorflow, and bugs which are related to regular programming mistakes like for any traditional software.\\
According to the categories in the most recent taxonomy of DL faults \cite{DL_faults}, we mention the type of faults that could be covered by the proposed rules: \textit{Wrong Tensor Shape, Wrong Shape of Input Data, Model Properties, Layers, and Loss Function}. Based on the count of manually-analyzed real-world buggy programs in~\cite{DL_faults}, we found that the covered DL bugs/issues in \tool{} represents $51.7\%$, of all reported DL buggy samples in the taxonomy. In the following, we report a prevalence ratio for each type of bugs, i.e., the number of buggy DL programs assigned to the underlying category divided by the total number of buggy DL programs in \cite{DL_faults}:
\begin{itemize}
\item \textit{Wrong Tensor Shape ($14.1\%$). }It refers to errors leading to unexpected tensor shape and mismatch between operations' shapes of tensors.
\item \textit{Wrong Shape of Input Data ($14.8\%$). }It assembles the bugs caused by invalid shapes of input data for a computational layer including input layer, hidden layers, and output layer, as well as the shape of math function's inputs.
\item \textit{Model Properties ($2.7\%$). }It comprises improper modeling choices that can dramatically degrade the DNN's performance such as missing, wrong model initialization, or sub-optimal model structure.
\item \textit{Layers ($15.4\%$). }It contains the bugs related to layers including missing, redundant, misconfigured and wrong neural network layers.
\item \textit{Loss Function ($4.7\%$). }It covers different issues in relation with the loss component such as missing, inadequate and wrong loss function.
\end{itemize}

\subsection{Representing Rules as Graph Transformations}
In this paper, the meta-model is presented as a type graph and each model is a graph, instantiating the type graph. Each DL program is converted to a graph, as well. As a straightforward approach, graph transformations are chosen to implement the verification rules. Each verification rule is implemented as one or some graph transformations or graph processing operators. In fact, graph transformations are used to detect possible faults in a model, faults that are caused by violating the verification rules. Consequently, a transformation is applicable where the conditions of the corresponding rule are violated. In other words, if conditions of a verification rule are violated representing a fault in a model then the graph operation(s) of that rule will be applicable. Graph transformations are very flexible to find violation of some conditions in a graph. Recalling that a graph transformation \textit{r} is defined by a triplet of \textit{$(LHS_r, RHS_r, NAC_r)$}, a specific condition would be checked by finding a match of \textit{$LHS_r$} in the graph and/or the absence of \textit{$NAC_r$}. Once a graph operation is applied, i.e., detecting a fault in a part of the graph, a specific fault code is added to the node or edge in which the violation occurred. This action is represented by the right hand-side of the rule \textit{$RHS_r$}.

\begin{figure}[t]
  \includegraphics[width=0.9\linewidth]{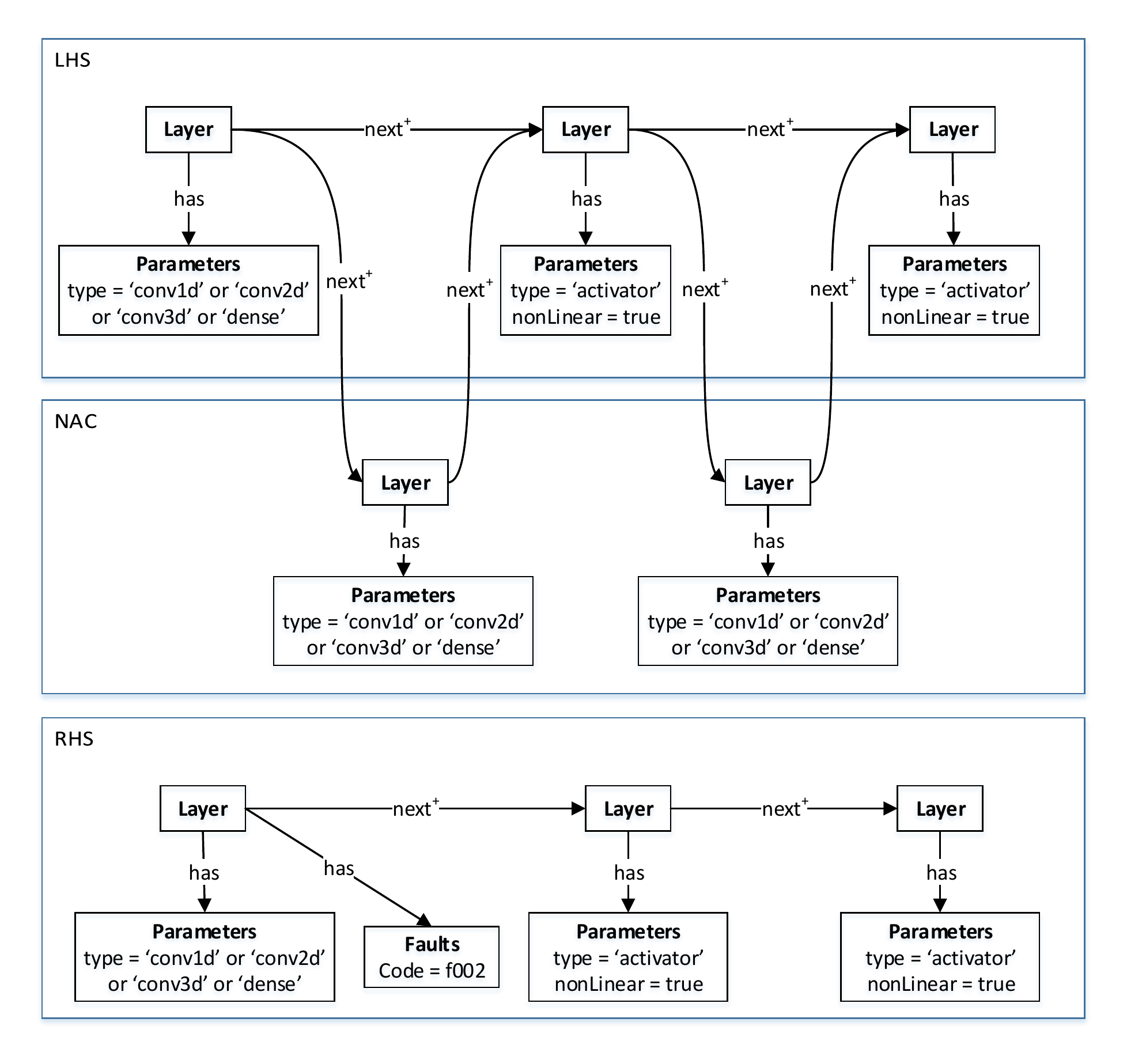}
  \vspace{-15pt}
  \caption{An Example of Graph Transformation Rules: Implementation of Rule 4.}
  \label{fig:rulegraph1}
  \vspace{-20pt}
\end{figure}

Figure \ref{fig:rulegraph1} illustrates one of verification rules implemented as a graph transformation, showing LHS, RHS, and NAC. The transformation is an implementation of Rule 4 which asserts that: \emph{Multiple and redundant connected activations are not allowed}. Developers usually add activations after learning layers (like convolution and dense layers) to produce proper output signals. LHS shows a learning layer with the \textbf{type} of `dense', `conv1d', `conv2d' or `conv3d' in its \textbf{Parameters} node, followed by two consecutive layers containing the \textbf{type} of `activator' and the \textbf{nonLinear} as \textit{True} in their parameters. A positive closure is used on the label of incoming edges to activation layers (\textit{next+}). This states that activations may appear in any \textbf{Layer} node on the path beginning from the learning layer and including multiple \textit{next} edges ($\geq 1$). To be sure that another learning layer would not appear on this path, e.g., false detection of the next learning layer followed by its only activation, NAC forbids the existence of any learning \textbf{Layer} node on the subpath leading to activation nodes. If such a match is found in a model (graph of DL program), Rule 8 is violated. Therefore, RHS just adds a \textbf{Faults} node with relevant fault code to the faulty component, i.e., learning layer. Because of space limitation, we cannot present in the paper all the graph transformations implemented for our model verification. We refer interested readers to the source code of \tool{} which is available online \cite{neuralint}.

\section{A Model-based Verification Approach for DL programs}\label{approach}
In this section, we describe our approach, \tool{} for detecting faults in DL programs. \tool{} is a model-based automated approach that performs a static analysis of a DL program to detect faults and design inefficiencies. Algorithm \ref{algo:algo1} shows the pseudocode of \tool{}. The inputs are a DL program and a graph grammar, i.e., a set of graph transformations rules. As presented in Algorithm \ref{algo:algo1}, \tool{} has three main steps: extract a graph from the DL program, perform graph checking and generate a report from the resultant graph. At first, the DL program is modeled as a graph that conforms to the proposed meta-model, i.e., type graph. Then, a checking process runs to find bugs/issues in the model. This process attempts to apply rules to the graph and stops when further rule application becomes impossible. Then, \tool{} traverses this graph to generate a report for the user, containing a description of the faults and design issues found for each component. Except graph checking and graph transformations, all other parts of \tool{} are implemented in Python. We discuss details of each step in the rest of this section.
The source code, developer’s guide and some examples are available online \cite{neuralint}.

\subsection{Modeling DL Program as Graph}
In our graph-based approach, a DL program is modeled by a graph instance conformed to the type graph, i.e., meta-model. To fulfill this primary step, we implement the graph generation relying on static code analysis that examines the source code and extracts the valuable code units and segments the information needed to instantiate the type graph's components. This analysis is performed without executing programs to extract the structure of the DL model from the code. Based on the OMG (Object Management Group) taxonomy of software analysis types~\cite{omg_static}, our graph generation process belongs to the category of technology-level software analysis because the code inspection routines are further customized to consider the interactions between identified units and connect them to the internals of the used technology (DL library). This provides a more holistic and semantic view of the analyzed DL program, that allows detecting faults related to either DL library's API misuse or DL algorithm implementation requirements. Hence, a specific graph generator should be implemented for each supported DL library. Without loss of its generality, \tool{} currently supports DL programs written using TensorFlow and Keras as two well-known and popular libraries. It should be noted that \tool{} can be extended to detect bugs/issues in DL programs developed by other DL libraries (like PyTorch), as well. The only necessary step is extending the parser to cover specific APIs of each DL framework. Moreover, by employing static analysis we are limited to the information available prior to the runtime, in contrast to dynamic analysis performed on programs while they are executing. Hence, we cannot detect bugs/issues depending on information introduced in the runtime environment, like dynamic types or dynamic constructions. However, we believe that the current approach can detect a significant amount of bugs/issues since DL programming is usually simple, and much useful information on DL models can be extracted by static analysis. In the following, we describe steps of modeling of DL programs as attributed graphs.\\
Our approach consists in parsing the DL routines called in the DL program line by line to extract the components related to both the DNN model and the DL training algorithm, as well as their configurations. The identified components are independent of the context of the parsed program. We used Abstract Syntax Tree (AST) to parse the DNN program script. AST represents the abstract synthetic structure of the scripts as a tree. This tree represents the abstract syntactic structure of the source code of the DL program. Each node of the tree denotes a construct or statement in the source code of a DL program. Arguments of function calls or assignment statements are extracted and stored in subtrees of the node. As we process the code line by line according to AST, the graph is constructed gradually by appending nodes and edges. In each DL library, there are specific built-in routines for defining various layers, configuring them (e.g., adding dropout and activations), connecting layers to each other, feeding input, calculating output, and training the network. In this way, the most important parts in the DL program for constructing its model like layers, dimensions, loss and optimizer functions could be identified by the parser. Algorithm \ref{algo:algo2} illustrates this process. Based on the information extracted by AST, for any line in the code of a DL program that indicates an assignment statement, we build a dictionary to store its value for further usage. For API calls that add or configure a layer, we firstly extract the related properties like type and number of units (neurons) from AST or dictionary of variables. Since the current version of \tool{} is designed to handle DL programs developed using TensorFlow or Keras libraries, we have covered the API of these frameworks. Sometimes, a computation phase is required for each layer to process its attributes and attach it to other layers in the graph correctly. For instance, the dynamic shapes of the layer's tensors (i.e., input and output data layer) should be computed. In such cases, the required properties are computed before adding the node to the graph. If an API call relates to compiling or training the model (building the DNN, adding loss and optimizer functions), the optimizer and loss nodes are added to the graph. At the end, we will check all the interconnected layers to verify the coherence of the datashapes flowing throughout the DNN’s computational layers and apply the required corrections. Afterward, the generated graph, including all relevant components (nodes/edges and their properties) based on the extracted and computed information is returned.

\begin{algorithm}[t]
\KwIn{A DL program, \textit{program}, and \textit{rules} as a graph grammar}
\KwOut{List of bugs or warnings to improve the program}
\SetAlgoLined
 $graph \leftarrow{} \text{extractGraphFromProgram(\textit{program})}$ (Algorithm \ref{algo:algo2})\\
 $final \leftarrow{} \text{graphChecker(\textit{graph}, \textit{rules})}:$\\
 \begin{enumerate}
  \item starting by \textit{graph}, apply enables rules.
  \item apply enabled rules recursively.
  \item terminate when further application of rules becomes impossible.
  \item \Return \textit{final}.
 \end{enumerate}
 $report \leftarrow{} \text{extractReportFromGraph(\textit{final})}$\\
 \Return \textit{report}
 \caption{\tool}
 \label{algo:algo1}
\end{algorithm}
\begin{algorithm}[t]
\KwIn{A DL program in Python developed by TensorFlow or Keras}
\KwOut{A graph indicating program's DL model with respect to the meta-model}
\SetAlgoLined
 $graph \leftarrow{} \text{empty}$\\
 $dictionary \leftarrow{} \text{empty}$\\
 \For{each line of DL program}{
    \If{line encodes an assignment statement}{
        extract left-hand (variable) and right-hand side (value) of the statement\\
        add the statement to the $dictionary$
    }
    \If{line encodes a layer}{
        extract properties of layer (type, size, kernel, padding, ...)
        \\look up values of variables in $dictionary$
        \\add the corresponding node(s) and edge(s) to the $graph$
    }
    \If{line encodes compilation of the model}{
        extract properties of compilation (loss and optimizer function)
        \\look up values of variables in $dictionary$
        \\add the corresponding node(s) and edge(s) to the $graph$
    }
    }
 \Return {\textit{graph}}
 \caption{Extracting graph from DL program}
 \label{algo:algo2}
\end{algorithm}
\subsection{Model-based Verification using Graph Transformations}
The verification rules are implemented as graph transformations to process and verify the graph. Each graph transformation applies to the graph if conditions of the rule are violated. Once the DL source code is modeled as a graph, the violations of rules can be detected with a graph transformation tool that executes the sequence of rules over the model of the DL program. In this paper, we have used the GROOVE toolset \cite{rensink2004groove} to perform graph operations. GROOVE is a tool for implementing, simulating, and analysis of graph transformation systems. It is capable of exploring recursively and collecting all possible rule applications over a host (start) graph. This is referred to as the exploration of the state space of a graph grammar. GROOVE explores the state space by applying a slightly modified version of standard graph traversal algorithms, like depth-first search (DFS) or breadth-first search (BFS). Furthermore, it has a graphical interface for editing graphs and rules, and for exploring and visualising the GTS which could be called via command line, as well. The output of GROOVE is called the final graph on which no further rule application is possible. For more information about GROOVE’s internal mechanism and its capabilities for modeling and simulating GTS, the interested reader may refer to \cite{ghamarian2012modelling}.\\
In order to find which rules are violated, the graph transformation system must be simulated. The simulation performed by GROOVE automatically applies the matching transformation rules over the graph of the DL program. Actually, this process generates a state space, in which the model of the DL source code (\textit{graph}) is the start state and the transitions are the applied transformation rules. It explores the state space of all graphs that are reachable from \textit{graph}. In certain states, no more transformation rules can be applied; these states are called final states. A path starting from the start state and leading to a final state, consists of applied transformations indicating the detected faults or violation of verification rules. Moreover, this path indicates the type and location of detected faults. A specific code for each type of fault has been associated with the faulty component when the rule has rewritten the graph.\\
The rules are implemented in such a way that starts from the first layer and proceeds to the next layers one by one. At first, the general structure and connectivity of deep neural networks is tested assuring that input, hidden and output layers are well-formed and connected. These transformation rules mark the graph components (nodes and edges) with relevant flags to indicate the performed tests. Then, each graph operation checks specific conditions that are asserted in its rule using the information provided in the graph. A transformation should be fired if a rule violation is observed in the model of a DL program. If there are multiple rule violations or various instances of a violation in the considered model, all of them will be detected by applying multiple enabled rules. At last, a parser is developed to process the final graph and extract information about detected issues/bugs to generate a report for the user.\\
\section{Empirical Evaluation}\label{evaluation} 
In this section, we report an empirical evaluation that aimed to assess the effectiveness of \tool{}.
\subsection{Studied Programs}\label{StudiedPrograms}
We have evaluated the effectiveness and efficiency of \tool{} in detecting bugs/issues on a set of synthetic and real-world faulty DL programs. To create realistic synthetic examples, we also need some real-world DL programs to imitate the faults occurring in them. To find a proper set of real-world faulty DL programs, we have used two main sources: 1) samples found by directly searching over SO with keywords related to the categories of bugs covered by \tool{}, and 2) public datasets of buggy DL programs (from SO and GitHub) released by previous research studies. For the former, we chose SO because it is the most popular Q\&A forum for software development. As of May 2020, it has collected more than $19$ million questions and $29$ million answers. It has been also leveraged by previous studies on DL software systems~\cite{DL_bugs_1, DL_bugs_2, DL_challenges}. Since \tool{} currently supports both TensorFlow and Keras, we searched SO posts tagged by one of these libraries with the objective of collecting buggy DL code including multi-granularity levels such as a single function call, a snippet (few lines) of code or a whole DL program. In fact, we found that SO assembles 57,104 and 27,008 questions, tagged respectively with TensorFlow and Keras, that comprise diverse issues encountered by DL practitioners when dealing with these libraries. Hence, we refined our search queries with keywords related to the categories of bugs covered by \tool{} that are described in Section \ref{rules} resulting in 255 posts. We manually inspected, for each type of bug, the top-10 relevant SO posts (i.e., according to built-in SO relevance criterion) mentioning one or more of its associated keywords. We consider SO posts, containing full code script or code snippets that are related to one or multiple bugs belonging to the above-mentioned categories. This process left us with 18 faulty DL programs.\\
Regarding public datasets of buggy DL programs (from SO and GitHub), we consider three publicly available datasets/replication packages \cite{DL_bugs_1, DL_bugs_2, DL_faults}. Also we consider another dataset from a recent research on bug fix patterns in DL programs developed by five popular DL libraries including Tensorflow and Keras \cite{DL_fix2020}. They have studied several repair patterns in DL programs. All these studies investigated various faulty DL programs from SO and GitHub. We have manually inspected all artifacts they have used in their study to find relevant faulty examples to evaluate \tool{}. Actually, finding proper samples for evaluating our tool is not an easy task. We explain the methodology followed and encountered difficulties in the rest of this section. Some DL programs were developed by libraries other than Tensorflow/Keras which are out of scope of the current version of \tool{}. In total, we had 733 SO posts and 682 samples from GitHub from all these sources where 622 programs were developed by TensorFlow and 793 by Keras. Among DL programs developed by Tensorflow/Keras, we have excluded programs containing types of faults that are not covered by the current version of \tool{}, for example those related to recurrent neural networks. So, we were left with 566 faulty DL programs. In the next round, programs developed with older versions of Tensorflow/Keras were discarded if the API related to the fault was not supported in later versions. Many of the remaining samples (89 from SO and 126 from GitHub), especially those from GitHub, were actually libraries (not DL programs) that have been developed on the top of DL libraries for particular problems or domains, e.g. image/speech processing, reinforcement learning, or natural language processing. We also discarded these libraries which were 86. Although their implementation contained bugs that led to buggy DL models, when the libraries were used to build DL models, we discarded them because they do not build a model explicitly using Tensorflow/Keras APIs. For example, they get a specific configuration file or a code written by their own high-level APIs as input, and use it to construct a DL model. Therefore, it is impossible to use those examples since the scope of \tool{} is defined to cover DL programs developed directly by employing Tensorflow/Keras built-in APIs. It should be noted that customized parsers can be developed to extract DL models from any configuration file or high-level code that is not currently covered by \tool{} and then use our tool to find bugs/issues in them. After processing all these artifacts, we ended up with 26 buggy DL programs shared on SO (18 from our direct search and 8 from public datasets) and 8 from GitHub.\\
Since \tool{} requires a full DL program to construct a graph on which the model verification is performed, we decided to prepare synthetic examples by a mix of synthetic code and reproduced real DL programs for the evaluation of \tool{}. The reproduction of buggy DL programs from the SO posts is quite difficult when a major part of the code is not provided in the post. Anyway, to reproduce real buggy DL programs, we proceed as follows: (1) we first implement two well-known CNN applications, LeNet \cite{lecun2015lenet} on MNIST data \cite{lecun1998mnist} and VGG-16 \cite{VGGNet} on Imagenet data \cite{deng2009imagenet}, as base programs. To enhance diversity at technology level as well, we use both of our supported DL libraries, Tensorflow for LeNet and Keras for VGG-16 following, respectively, the official implementations \cite{lenet_TF} and \cite{VGG_Keras} published on GitHub; (2) regarding implementation-related bugs, we inject each fault found, to one of the base DL programs; (3) regarding design-related issues, we poorly re-designed the structure of the base program's model to include inefficiencies violating the common patterns and best practices mentioned in Section \ref{rules}. Finally, we constructed a total of 28 buggy synthetic programs which corresponds to one or two examples per detection rule. We constructed two faulty examples for a rule when there are two contexts in which the rule can be triggered, one example for each of these contexts. For instance, Rule 10, which validates the loss linkage, has been evaluated against both contexts of binary cross-entropy (used for binary class problem) and categorical cross-entropy (used for multi-class problem). For injecting bugs, we followed fault patterns observed in real buggy samples during our rule extraction process (illustrated on Figure \ref{fig:methodology}). Hence, the injected bugs are realistic reproductions of faults. Our goal for evaluating \tool{} using synthetic examples is debugging, i.e., making sure of its accuracy and effectiveness prior to evaluating it on real-world examples. For more details, please see our replication package containing all samples and implemented synthetic code \cite{neuralint}.\\
Based on DL bug symptoms defined in \cite{DL_bugs_2}, we found three bug symptoms in our studied DL programs: \textit{i) Bad performance. }Bad or low performance is a common effect of conceptual issues related to design structure inefficiency or poor choices, misconfiguration of DL components; \textit{ii) Incorrect Functionality. }This symptom refers to situations where the DL program behaves in an unexpected way without any runtime or compilation error. For instance, the DNN outputs only one label among class labels; \textit{iii) Program Crash: } This bug effect is common for all software programs, it means that the program stopped running and raised an exception. Regarding the recommended fixes, we examined the accepted or endorsed answers of SO users to determine the bug-fixing repair (i.e., accepted) or recommendations provided to guide the user who asked the question towards finding the root cause of the error (i.e., open question).

\subsection{Results}
First, we have evaluated \tool{} using 28 synthesized examples to investigate the correctness and preliminary effectiveness of the proposed approach. \tool{} has successfully detected the bugs and issues in all synthetic examples. It should be noted that \tool{} extracts the DL model from a DL program and employs graph transformations to apply the proposed rules on the model, not the code, to detect possible bugs/issues. For extracting the model from the code, \tool{} relies on TensorFlow/Keras APIs and not on any particular patterns in the code. Moreover, we have not limited our experiments to these synthetic samples and have tested \tool{} on real-world faulty DL programs. In other words, the tool is evaluated on faulty DL programs with bug patterns that were not considered when creating the tool or synthetic examples.\\
To evaluate practical effectiveness and accuracy of \tool{}, 34 real-world DL programs from SO posts and GitHub repositories are used. Results are presented in Table \ref{table:table-results-1} and Table \ref{table:table-results-3}. Table \ref{table:table-results-1} reports results over DL programs extracted from SO posts. For each DL program, we report the ID of the post over SO, reported symptoms of buggy programs by the developer, fixes recommended by other users, output of \tool{} (violated rules), number of true positive cases and false negative cases respectively. True positives are reported as a+b where a is the number of bugs/issues reported by SO users that are detected by \tool{}, and b is the number of bugs/issues detected by \tool{} that are not mentioned by SO users. For b, two of the authors independently have checked each program and the output of \tool{} manually to ensure that the output is correct. The total number of false positive cases is zero, so we do not report them. It is well-known that the best practice is analyzing SO posts with accepted answers (No. 1 to 20 in Table \ref{table:table-results-1}) ensuring the proposed solution is a real fix and addresses the mentioned problem. In our searching process for faulty samples, however, we have encountered 6 posts in SO without accepted answers (No. 21 to 26 in Table \ref{table:table-results-1}) containing relevant DL buggy programs or code snippets. Although none of the provided answers in these posts were accepted by the user who asked the question, we found at least one helpful and correct answer in the posts after a careful analysis. Specifically, one of the authors has manually inspected answers to make sure that SO users pointed out a right solution to the problem according to our verification rules. This process has been verified by another author assuring that we have a correct assessment and that the output of \tool{} is accurate. 
Regarding samples from GitHub, the results are reported in Table \ref{table:table-results-3}. True positives are again reported as a+b where a is the number of bugs/issues that are successfully detected by \tool{} according to reported problems in GitHub or a previous research study as mentioned in Subsection \ref{StudiedPrograms}. On the other hand, b is the number of bugs/issues detected by \tool{} but not reported in GitHub or a previous study. Similar to what we have done for SO posts, we have checked each program and the output of \tool{} manually to ensure that the output is correct. In all tables, the rules which detect the bug/issue as reported by the developer are highlighted in bold letters. Detailed information of each sample including the link to GitHub repositories are available in our replication package \cite{neuralint}.\begin{table*}
\caption{Results of validating \tool{} using real DL programs selected from StackOverflow.} 

\vspace{-5pt}
\label{resultsTable}
\resizebox{.99\textwidth}{!}{
\begin{tabular}{|c|c|p{2.7cm}|p{5.7cm}|p{2.1cm}|c|c|}
 \hline
 No.&SO \#&Symptom & Recommended Fix& Violated Rules & TP & FN\\
 \hline
 \hline
 1 & 44399299 & Program Crash & Change the shape of the input layer&\textbf{7}&1+0&0\\
 \hline
 2 & 43464835 & Program Crash & Change the shape of the input layer&-&0+0&1\\
 \hline
 3 & 42913869 & Program Crash &  Change the number of units for the output layer &  3&0+1&1\\
 \hline
 4 & 48518434 & Program Crash & Reduce spatial size of both Conv. filtering and pooling widows &  \textbf{7} & 1+0 & 0\\
 \hline
 5 & 40857445 & Program Crash & Adding a flatten layer & \textbf{6}&1+0&0\\
 \hline
 6 & 50555434 & Bad Performance & Use softmax activation instead of sigmoid and categorical\_crossentropy loss instead MAE & \textbf{10}&1+0&0\\
 \hline
 7 & 46177505 & Program Crash & Change spatial size of Conv. filtering and pooling widows & 5, 10&0+2&1\\
 \hline
 8 & 50426349 & Program Crash & Change the shape of the input layer &  19, 20 & 0+2&1\\
 \hline
 9 &38584268 & Program Crash &  Adding a flatten layer& \textbf{6}, 21&1+1&0\\
 \hline
 10 & 45120429  & Program crash & Change the number of units for the output layer, Adding a flatten layer& \textbf{6, 19}, 10&2+1&0\\
 \hline
 11 & 45378493  & Incorrect Functionality &Use a sigmoid for last layer activation& \textbf{5, 10}, 16, 19, 20&2+3&0\\
 \hline
 12 &45711636 & Program Crash & Use channels\_last format for input data &  \textbf{7} & 1+0&0\\
 \hline
 13 & 34311586 & Bad Performance & Remove the last layer activation& \textbf{5, 10}, 19&2+1&0\\
 \hline
 14 & 50079585\_1  & Bad Performance &Use softmax activation instead of sigmoid and categorical\_crossentropy loss instead binary\_crossentropy &-&0+0&1\\
 \hline
 15 & 50079585\_2  & Incorrect Functionality &Change the number of units for the output layer& \textbf{10}&1+0&1\\
 \hline
 16 & 51749207 & Bad Performance &Use of sigmoid activation instead of softmax &  \textbf{5, 10}, 19&2+1&0\\
 \hline
 17 & 53119432  & Program Crash & Adding a flatten layer& \textbf{6}, 19&1+1&0\\
 \hline
 18 & 55731589 & Program Crash & Use of 'same' instead of 'valid' for layer padding type&  \textbf{7} & 1+0 & 0\\
 \hline
 19 & 58844149 & Bad Performance & Use of sigmoid as last layer activation&  \textbf{5, 10}, 21& 2+1 & 0\\
 \hline
 20 & 61030068 & Program Crash & Adding a flatten layer&  \textbf{6}& 1+0 & 0\\
 \hline
 21 & 33969059 & Bad Performance & Change the number of units for the output layer&\textbf{10} & 1+0 & 0\\
 \hline
 22 &44184091 & Program Crash & Fix the limit size for input sequence data&15 & 0+1 & 1\\
 \hline
 23 &44322611 & Bad Performance & Prune the DNN, use RMSprop instead SGD &10, 20, 21& 0+2 & 1\\
 \hline
 24 & 49117607& Program Crash & Reduce spatial size of both Conv. filtering and pooling widows & 16& 0+1 &0\\
 \hline
 25 & 55776436 & Bad Performance & Try Data augmentation, Regularization, filtering spatial size reduction, and DNN Depth Increase& \textbf{7, 16, 17, 20} & 4+0 & 0\\
 \hline
 26 & 60566498 & Bad Performance & Try Data augmentation and Hyperparameters Tuning  & \textbf{15}, 16 & 1+1 & 0\\
 \hline
 \end{tabular}
\label{table:table-results-1}
}
\vspace{-5pt}
\end{table*}\\
In total, 31 out of 44 bugs/issues are detected correctly by \tool{}, so the recall is evaluated as 70.5 \%. All of these bugs/issues were identified by users/developers or previous research studies. The recall for SO posts with accepted answers is 76.9 \% (20 out of 26), for SO posts without accepted answers is 75 \% (6 out of 8), and for GitHub samples is 50 \% (5 out of 10). The precision is 100 \% meaning that we do not observe any false positive case in our evaluation. Moreover, \tool{} correctly detected 33 additional bugs/issues that were not reported by users/developers who commented on the SO posts or GitHub repositories. Most of them, 29 out of 33, are design issues. \tool{} has successfully detected 64 bugs/issues in 34 real DL programs in overall.\\
While fewer bugs are detected by \tool{} in GitHub samples compared to SO posts, more design inefficiencies are detected by \tool{} in GitHub samples (14 in 8 samples). Also, since \tool{} is based on a static analysis, being able to detect half of the faults contained in the studied Github projects is already an interesting feat, since it allows developers to catch them early on, before they have to run their programs. Based on these results, we can report that the performance of \tool{} is noteworthy; about three-quarters of known bugs are successfully detected (recall) and a significant number of hidden bugs and design inefficiencies of DL programs. Moreover, its precision is 100 \% meaning that while the tool may miss some faults in the evaluated DL programs (overall recall is 70.5 \%), it never detects bugs/issues wrongly. The reason is that our detection process is based on proposed verification rules and we report their violations in DL models extracted from DL programs.\\
We have performed the experiments using a machine with Intel i7-9750H CPU and 16GB of main memory running Windows 10. The average execution time of \tool{} for the studied TensorFlow and Keras samples are 1.800 and 2.049 seconds, respectively. It should be noted that graph checking (performed by GROOVE) consumes the main portion of the execution time, about 99.7 \%. Our preliminary analysis revealed that the running time mainly depends on the number of layers of the DL model. Details of the execution time of \tool{} for five real DL programs with different sizes are reported in Table \ref{timeTable}. According to these results, the execution time of extracting and checking the graph increases as the number of layers grows. Extracting the graph is accomplished by single or multiple passes through the code. Hence, the execution time grows linearly by the number of layers as adding/configuring each layer needs a few API calls in TensorFlow/Keras. On the other hand, GROOVE supports priority-based rule application as well as various search strategies to explore the full state space, i.e., checking and applying all applicable rules in each state \cite{ghamarian2012modelling}. We have used BFS and priority-based rule application to improve the efficiency. However, the execution time of graph checking grows faster than extracting the graph as the number of layers of the DL model increases. The running time of \tool{} can be improved which is left for future work. 
\newcounter{rownumbers}
\newcommand\rownumber{\stepcounter{rownumbers}\arabic{rownumbers}}
\begin{table*}
\caption{Results of validating \tool{} using real DL programs selected from GitHub.}
\vspace{-5pt}
\label{resultsTable}
\resizebox{.99\textwidth}{!}{
\begin{tabular}{|c|p{3cm}|p{6cm}|p{2.1cm}|c|c|}
 \hline
 No.& Symptom & Recommended Fix & Violated Rules & TP & FN\\
 \hline
 \hline
 \rownumber & Bad Performance \cite{github1} & Changing the last layer activation& 19, 20, 21&0+3&1\\
 \hline
 \rownumber & Bad Performance \cite{github2} & Changing layer dimensions & 19, 20, 21&0+3&1\\
 \hline
 \rownumber & Bad Performance \cite{github3}& Changing layer dimensions (padding)&\textbf{4}, 19, 21&1+1&1\\
 \hline
 \rownumber & Bad Performance \cite{github4}& Adding a pooling layer & \textbf{5}, 16, 20, 21 & 1+3 & 1\\
 \hline
 \rownumber & Bad Performance \cite{github5}& Changing layer dimensions & \textbf{19}, 16, 20, 21&1+3&0\\
 \hline
 \rownumber & Bad Performance \cite{github6}& Adding ReLU activation to the last layer &\textbf{3}&1+0&0\\
 \hline
 \rownumber & Bad Performance \cite{github7}& Adding ReLU activation to the last layer &\textbf{3}&1+0&0\\
 \hline
 \rownumber & Program Crash \cite{github8}& Changing layer dimensions & 19 & 0+1 & 1\\
 \hline
\end{tabular}
\label{table:table-results-3}
}
\vspace{-5pt}
\end{table*}

\subsection{Limitations and Discussion}
GitHub samples were developed by advanced developers and are more complex than SO posts. While the scope of \tool{} is defined to cover convolutional architecture as a particular type of FNN designed mainly for classification of 2D images, audio spectrograms, or 3D videos, some of studied GitHub samples have used convolutional architecture for data generation (e.g., extraction of structural lines of images\footnote{\url{https://github.com/hepesu/LineDistiller}}) or text classification\footnote{\url{https://github.com/bwallace/rationale-CNN}}. In another sample\footnote{\url{https://github.com/cmasch/densenet}}, developers concatenated outputs of multiple convolutional architecture, each layer taking all preceding feature-maps as input, which is not frequent in popular CNNs. Using popular convolutional architectures such as VGG, ResNet, or MobileNet as a part of a DL model or modifying them for particular tasks is also observed in studied samples from GitHub\footnote{\url{https://github.com/mateusz93/Car-recognition/commit/94b36ea}}. Although the proposed meta-model is capable of representing these models as FNNs, particular rules must be proposed to find faults and improve the accuracy of \tool{} on these samples. The developer added a dropout layer after dense layers to fix the problem in one of GitHub samples\footnote{\url{https://github.com/dishen12/keras_frcnn/commit/38413c6}}. Although, as mentioned in Subsection 2.2, regularization methods are required to improve the convergence and generalizability of DL models, we need more investigations for proposing a rule to detect lack of enough or proper regularizations.\\ 
The focus of the design and implementation of \tool{}, is on faults that relate to structural (architectural) properties of DL programs rather than their dynamic properties that need the programs to be executed. In other words, there are some frequent types of bugs/issues in DL programs that could be detected without dynamic analysis of the DL program \cite{DL_faults}. However, the lack of dynamic analysis of DL programs is a limitation of our approach. Such analysis would allow for the detection of runtime bugs and bugs/issues in training/inference of DL models. For example, in program No. 22 in Table \ref{table:table-results-1}, the mismatch shapes is caused by the size of loaded input size during the execution, it is a runtime bug and could not be detected by the current version of \tool{}. Some other bugs in DL programs need in-depth runtime analysis. For example, using dropout before batchnorm makes the behavior of DNN different during training and evaluation phases. This is the case for program No. 15 in Table \ref{table:table-results-1}. In another sample\footnote{\url{https://github.com/taashi-s/UNet_Keras/commit/b1b6d93}}, batch size has been modified to improve the performance of the learning phase which cannot be investigated without analyzing the learning performance during runtime. Detecting these faults is currently out of scope of \tool{} and to cover them, DL programs must be experimentally tested to assess their performance for the underlying problem and then detect it.\\
Another challenge that we faced is related to the multiple releases of TensorFlow library that significantly changed the API functions; which makes the graph generator mainly designed and implemented in regards to the 1.15 version, incapable of detecting some of the required components for versions other than 1.15.\\
Lack or limited access to real DL programs annotated with possible bugs, design inefficiencies and recommended fixes to evaluate DL testing approaches accurately and effectively could be regarded as a barrier in this line of research. Finally, the current version of \tool{} could find problems in FNNs, particularly CNNs. Other neural network architectures, like recurrent neural networks are out of the scope of this version. \tool{} could be applied to detect bugs/issues in such neural networks by extending the meta-model to capture their properties, i.e., possibility of connections between the nodes that form a cycle. Also, new rules could be defined to detect specific problems in each architecture according to frequent observed bugs/issues or best practices.
\begin{table*}
\caption{Execution time of \tool{} for five real DL programs with different sizes (times are in seconds).}
\vspace{-5pt}
\label{timeTable}
\resizebox{.6\textwidth}{!}{
\begin{tabular}{|c|c|c|c|}
\hline
\multirow{2}{*}{No.} & {\multirow{2}{*}{Number of layers}} & Running time of & Running time of\\
& & Graph extraction & Graph checking\\

\hline
\hline
1 & 6 & 0.003 & 1.757\\
\hline
2 & 8 & 0.003 & 1.787\\
\hline
3 & 12 & 0.002 & 1.836\\
\hline
4 & 13 & 0.003 & 1.906\\
\hline
5 & 38 & 0.004 & 3.111\\
\hline
\end{tabular}
}
\vspace{-5pt}
\end{table*}

\section{Related Works}\label{relatedWork}
The current work on model verification of DL programs using meta-modeling and graph transformations is related to a number of approaches in the literature.\\
\textbf{Meta-modeling and deep learning}: Perez has introduced the idea of meta-model for deep learning in \cite{perez20meta}. After sketching a model for deep learning by noting UML meta-modeling, he argued that entities like Layers, Objectives, Activations, Optimizers, Metrics in high-level DL libraries are meta-models for deep learning. However, the paper keeps its interesting ideas in a high-level of abstraction. A meta-model for meta-learning is presented in \cite{hartmann2019meta}, alongside a discussion of its possible integration into state-of-the-art modelling frameworks. Meta-learning or automatic machine learning is a recent approach that applies learning algorithms on metadata about machine learning experiments to automatically improve the learning process. The concept of meta-learning is viewed from a meta-modeling point of view and its variabilities are described. Although the presented meta-model has similarities with our meta-model, the scope is different: while they focused on meta-learning, in this paper, we present a meta-model of DL programs. Moreover, \cite{hartmann2019meta} did not explore the possibility of identifying bugs in learning programs through model verification.\\
\textbf{Graph representation and transformations for program verification}:  Ciraci et al. have proposed a graph-based automatic process for verification of static program constraints on elements like macros and comments \cite{ciraci2010graph}. Their main contribution is introducing a meta-model which is called Source Code Modeling Language (SCML) to express constraints over program elements. \tool{} also employs meta-modelling and a graph-based approach while it focuses on detecting bugs and bad practices in DL programs. Hoppity is a tool to detect and fix bugs in Javascript programs \cite{hoppity2020}. The researchers formulate the problem as learning some consecutive graph transformations. The buggy program is modelled by a graph structure, and then Hoppity predicts the location and type of bugs and some graph edits to fix the detected bugs. Hoppity employs the abstract syntax tree (AST) to model the buggy program as a graph while we have used a meta-model and then extract relevant information from AST to build the model. Moreover, we propose the graph transformations for localization and detection of bugs in \tool{} without adoption of a learning algorithm. Researchers have proposed other approaches \cite{iyer2020software, allamanis2018learning} that have learned graphs from the source code of programs (as proper representations of program semantics) for bug detection. However, in \tool{} we do not use learning algorithms to find or configure graphs/graph transformations. \\
\textbf{Testing Different Aspects of DL Software Systems}: Over the last few years, researchers have developed multiple testing approaches for DL software systems~\cite{rev_ml_testing}. While the larger part of these approaches test the exactness of the model predictions against instances from the input data distribution~\cite{deepxplore, deeptest, deepevolution}, a few other approaches~\cite{metmorphic_testing, proof_ML, tfcheck} focused on the validation that DL training programs are  bug-free. Dwarakanath et al.~\cite{metmorphic_testing} have proposed high-level metamorphic relations that can detect bugs altering substantially the program behavior, but they do not provide guidance to identify the root causes and have not studied real buggy DL software. In contrast, \tool{} is able to detect common DL software bugs and to steer the developer towards identifying and fixing the root cause. Selsam et al.~\cite{proof_ML} leverage machine-checkable proofs to validate the math consistency of the DNN training program. However, a proof-based approach could be difficult to adopt in practice because it requires code written using low-level mathematical libraries, and nowadays, DL practitioners mostly use third-party DL libraries to construct reliable and scalable DL software systems. On the contrary, \tool{} is quite extensible and can cover any API by constructing a new graph generator dedicated to the target library. Braiek and Khomh~\cite{tfcheck} proposed, TFCheck, a property-based testing method for DNN training programs, which continuously collects metrics on training dynamics and performs rule-based assertions to validate that a DNN program behaves properly. Nevertheless, it requires a running DNN training program, which is unfortunately not always available. For instance, an error of mismatched shapes leads the graph construction to fail and cause the crash of the DNN program. To fill-in the gap, \tool{} detects issues in the DL software program through its written code without even a single run. 
\section{Conclusion}\label{conclusion}
In this paper, we have introduced \tool{}, a model-based verification approach for DL programs. As a model-driven approach, a meta-model has been proposed for DL programs at first. We have defined a set of verification rules for DL programs based on the meta-model. A model of each DL program is configured by parsing its code to extract relevant information. Afterward, a graph checking process is performed to verify the model and detect potential bugs or design inefficiencies. Graph transformation systems are used to implement the verification rules and modeling approach. The meta-model is represented by a type graph, DL programs are modeled as host graphs, and graph transformations execute the verification rules. \tool{} has been evaluated using synthetic and real DL programs. The results show that \tool{} effectively detects faults and design issues in both synthetic and real-world DL programs with a recall of 70.5 \% and a precision of 100 \%. Refining (or even redesigning) the meta-model and graph transformations are required to improve the accuracy and detect false negative cases. The proposed meta-model is designed for FNNs but can be extended to support other neural network architectures such as recurrent neural networks. Another direction of research for the future is expanding our set of verification rules to cover more types of issues occurring in DL programs. While \tool{} currently supports DL programs developed by TensorFlow/Keras, it can be extended to detect bugs/issues for other DL libraries (like PyTorch) by extending its parser to cover specific APIs of each DL library.

\begin{acks}
This work is partly funded by the Natural Sciences and Engineering Research Council of Canada (NSERC) and the Fonds de Recherche du Québec (FRQ).
\end{acks}

\bibliographystyle{ACM-Reference-Format}
\bibliography{sample-base}






\end{document}